\documentclass[aps,pra,titlepage,twocolumn,longbibliography]{revtex4-1}

\usepackage{amsmath}
\usepackage{amsfonts}
\usepackage{xcolor}
\usepackage{maybemath}
\usepackage{bm}
\usepackage{bbm}
\usepackage{graphicx}
\usepackage{dcolumn}
\usepackage{cancel}

\newcommand{\dd}{\mathrm{d}}
\newcommand{\ee}{\mathrm{e}}
\newcommand{\ii}{\mathrm{i}}

\def\v{\vec}

\newcommand{\calE}{\mathcal{E}}
\newcommand{\calO}{\mathcal{O}}

\newcommand{\rmTr}{{\text{Tr}}}

\newcommand{\trc}{\mathrm{trc}}

\definecolor{garrosgreen}{rgb}{0.1, 0.4, 0.1}
\definecolor{dartmouthgreen}{rgb}{0.05, 0.5, 0.06}
\definecolor{duelferred}{rgb}{0.7, 0.2, 0.1}
\definecolor{cambridgeblue}{rgb}{0.1, 0.3, 1.0}
\definecolor{oxfordblue}{rgb}{0.05, 0.2, 0.7}

\definecolor{light}{gray}{0.90}
\definecolor{darker}{gray}{0.50}
\definecolor{dark}{gray}{0.30}

\begin{document}

\title{Revisiting the Divergent
Multipole Expansion of Atom--Surface Interactions:\\
Hydrogen and Positronium, $\boldsymbol\alpha$--Quartz, 
and Physisorption}

\author{Ulrich D. Jentschura}
\affiliation{Department of Physics and LAMOR, Missouri University of Science and
Technology, Rolla, Missouri 65409, USA}

\begin{abstract}
We revisit the derivation of multipole contributions
to the atom-wall interaction 
previously presented in [G. \L{}ach {\em et al.}, 
Phys. Rev. A 81, 052507 (2010)].
A careful reconsideration of the 
angular-momentum decomposition of the 
second-, third- and fourth-rank tensors 
composed of the derivatives of the 
electric-field modes leads to a
modification for the results 
for the quadrupole, octupole and hexadecupole contributions 
to the atom-wall interaction.
Asymptotic results are given for the asymptotic 
long-range forms of the multipole terms,
in both the short-range and long-range limits.
Calculations are carried 
out for hydrogen and positronium in contact 
with $\alpha$-quartz; a reanalysis of analytic 
models of the dielectric function of $\alpha$-quartz
is performed. Analytic results
are provided for the multipole polarizabilities
of hydrogen and positronium.
The quadrupole correction is
shown to be numerically significant for atom-surface interactions.
The expansion into
multipoles is shown to constitute a divergent,
asymptotic series. Connections 
to van-der-Waals corrected density-functional theory
and applications to physisorption are described.
\end{abstract}

\maketitle

\tableofcontents

%
% Introduction
%
\section{Introduction}
\label{sec1}

Multipole corrections to the atom-surface 
interactions have been considered in 
Refs.~\cite{LaDKJe2010pra},
with an application to helium interacting
with $\alpha$-quartz.
The theory of atom-surface interactions is well
known, for both perfect conductors
as well as realistic dielectric 
materials (see Refs.~\cite{Li1955,LaLi1960vol8,DzLiPi1961spu,%
DzLiPi1961advphys,Pa1974,SpTi1993,TiSp1993a,TiSp1993b},
pp.~261--263 of Ref.~\cite{Mi1994},
and Ref.~\cite{LaDKJe2010pra}).
In Ref.~\cite{LaDKJe2010pra},
the quadrupole, octupole and hexadecupole contributions to the 
atom-wall interaction were evaluated.
Here, we update the 
analysis presented in Ref.~\cite{LaDKJe2010pra}
with an emphasis on the isolation 
of angular-momentum components of the 
quantized electric field 
near the wall. 

Let us discuss the basis 
for the update, in terms of the initial expressions
used for the evaluations, both for the 
cases discussed in Ref.~\cite{LaDKJe2010pra}
as well as for the modifications discussed here.
Specifically, according to 
Eq.~(22) of Ref.~\cite{LaDKJe2010pra}
(in accordance with Ref.~\cite{Mi1994}),
the dipole part of the atom-wall 
interaction can be expressed as follows,
\begin{equation}
\label{e1}
\mathcal{E}_{\ell = 1} =
-\frac{1}{2} \sum_{\vec{k}\lambda}
\sum_i \alpha_{\ell = 1}(\omega)
\big| E^i_{\vec{k} \lambda}(\vec r)\big|^2 \,,
\end{equation}
where $\vec{E}_{\vec{k} \lambda}(\vec r)$ is the 
(in general complex)
mode function of the electric field near the 
wall, $\vec k$ is the wave vector,
and $\lambda$ is the polarization.
The index $i$ enumerates the Cartesian components.
The dipole polarizability of the atom
is $\alpha_{\ell = 1}(\omega) = \alpha_1(\omega)$, and it 
is evaluated at the angular frequency 
$\omega = \omega_k = c \, | \vec k |$,
where $c$ is the speed of light.
The mode function $\vec{E}_{\vec{k} \lambda}(z)$ 
of the electric field is related to the 
mode function $\vec{A}_{\vec{k} \lambda}(z)$ 
of the vector potential by the relation
\begin{equation}
\vec{E}_{\vec{k} \lambda}(\vec r) = \ii \omega_k \, 
\vec{A}_{\vec{k} \lambda}(\vec r) \,,
\end{equation}
where the mode functions
in Eqs.~(18),~(19) and (20) of Ref.~\cite{LaDKJe2010pra}.
The formalism for the treatment of the 
dipole contribution is well established.

The starting point for the analysis of the 
quadrupole, octupole, and hexadecupole corrections to the 
atom-wall interaction has been 
given by Eq.~(8) of Ref.~\cite{LaDKJe2010pra}
as follows,
\begin{equation}
\label{EE2}
\mathcal{E}_{\ell = 2} =- \frac{1}{12}\, \sum_{\vec{k}\lambda}\sum_{ij}
\alpha_2(\omega)\,
\big\vert \nabla^j E^i_{\vec{k} \lambda}(\vec{r})\big\vert^2 \,,
\end{equation}
where $\nabla^j \equiv \partial/\partial x^j$ 
and the indices $i$ and $j$ stand for the Cartesian components,
and $\alpha_2$ is the quadrupole polarizability of the 
atom. 
We reserve the index $i$ for the component of the
electric field. 
The octupole energy shift has been analyzed
in Ref.~\cite{LaDKJe2010pra} based on the expression
\begin{equation}
\label{EE3}
\mathcal{E}_{\ell = 3} = -\frac{1}{180}
\sum_{\vec{k}\lambda}\sum_{ijk}
\alpha_3(\omega)\,
\big\vert \nabla^j \nabla^k E^i_{\vec{k} \lambda}(\v{r})\big\vert^2 \,,
\end{equation}
where $\alpha_3 = \alpha_{\ell = 3}$ is the octupole polarizability of the atom.
The expressions~\eqref{EE2} and~\eqref{EE3} 
need to be substantiated by the specification
of the angular-momentum component 
relevant to the analysis. Namely, 
the virtual atomic transitions contributing 
to the quadrupole, octupole and hexadecupole polarizabilities 
of the atom connect the atomic ground state
to virtual states with angular momenta $\ell = 2$ and $\ell = 3$.
However, when one calculates sums such as
\begin{equation}
\sum_{ij} \big\vert \nabla^j E^i_{\vec{k} \lambda}(\vec{r})\big\vert^2 \,,
\end{equation}
without separating the ($\ell = 2$)-component of the 
second-rank tensor $\nabla^j E^i$ first, then
one sums over the ($\ell = 0$)- and ($\ell = 1$)-components
of the second-rank tensor $T^{ji} = \nabla^j E^i$ 
as well. While the proper isolation 
of the angular-momentum components does not 
change the functional form of the results
reported in Ref.~\cite{LaDKJe2010pra},
some prefactors receive modifications.
For the quadrupole term, 
the proper isolation of the ($\ell = 2$)-component
leads to the modified expression
\begin{equation}
\label{EEL2}
\mathcal{E}_{\ell = 2} =- \frac{1}{12}\,
\sum_{\vec{k}\lambda}\sum_{ij}
\alpha_2(\omega)\,
\big\vert
\left. \{ \nabla^j E^i_{\vec{k} \lambda}(\vec{r})
\} \right|_{\ell = 2}  \big\vert^2 \,.
\end{equation}
where the ($\ell = 2$)-component of a tensor is 
given by the traceless, symmetric component as 
follows,
\begin{equation}
\left. T^{ij} \right|_{\ell = 2} =
\frac12 \, \left( T^{ij} + T^{ji} \right) -
\frac13 \, \delta^{ij} \, \rmTr(T) \,,
\end{equation}
$\delta^{ij}$ is the Kronecker delta, and
$\rmTr(T) = \sum_i T^{ii}$ is the trace.
The octupole term is substantiated by the 
expression
\begin{equation}
\label{EEL3}
\mathcal{E}_3 = -\frac{1}{180}
\sum_{\vec{k}\lambda}\sum_{ijk}
\alpha_3(\omega)\,
\big\vert
\left. \{ \nabla^j \nabla^k E^i_{\vec{k} \lambda}(\vec{r}) \}
\right|_{\ell = 3}  \big\vert^2 \,,
\end{equation}
where the isolation of the octupole 
($\ell = 3$)-component of a third-rank tensor 
has recently been discussed by Itin and Reches in Ref.~\cite{ItRe2022}.
The isolation of the 
($\ell = 3$)-component of a third-rank tensor 
is not completely trivial.
The modified expressions given in Eqs.~\eqref{EEL2}
and~\eqref{EEL3} have phenomenological consequences 
which are discussed in the following.

Finally, the hexadecupole energy shift 
is calculated as follows,
\begin{equation}
\label{EEL4}
\mathcal{E}_4 = -\frac{1}{5040}
\sum_{\vec{k}\lambda}\sum_{ijk}
\alpha_4(\omega)\,
\big\vert
\left. \{ \nabla^j \nabla^k \nabla^l E^i_{\vec{k} \lambda}(\vec{r}) \}
\right|_{\ell = 4}  \big\vert^2 \,,
\end{equation}
The isolation of the ($\ell = 4$)-component from a Cartesian
tensor is discussed in Ref.~\cite{AnGh1982}.

We use Syst\'{e}me International (SI mksA) units in the following,
before we switch to atomic units in Sec.~\ref{sec4}.
The basis for the treatment of
multipole interactions is discussed in Sec.~\ref{sec2},
the case of a perfect conductor in Sec.~\ref{sec2B},
and dielectric materials in Sec.~\ref{sec3}.
Interactions with hydrogen and positronium atoms
are discussed in Sec.~\ref{sec4}.
Conclusions are drawn in Sec.~\ref{sec5}.

%
% Multipole Interactions
%
\section{Multipole Interactions}
\label{sec2}

%
% General Formulation
%
\subsection{General Formulation}

In order to put things into perspective, 
and in terms of some orientation,
we briefly review the 
concept of the multipole oscillator strength,
inspired by Chap.~5 of Ref.~\cite{JeAd2022book}.
The $2^\ell$-pole polarizability is 
written as 
\begin{multline} 
\label{multipole}
\alpha_\ell(\omega) = 
\frac{1}{2 \ell + 1} \sum_m
\left< 1S \left| Q_{\ell m} 
\left( \frac{1}{H - E_{1S} + \hbar \omega} 
\right.  \right.  \right.
\\
+ \left. \left. \left. \frac{1}{H - E_{1S} - \hbar \omega} \right)
Q^*_{\ell m} \right| 1S \right> \,,
\end{multline} 
where the $Q_{\ell m}$ tensor is given as
\begin{equation}
\label{QLm}
Q_{\ell m} = \sum_a e\, \sqrt{\frac{4 \pi}{2 \ell + 1}} \, 
r_a \, Y_{\ell m}(\hat{r}_a) \,,
\end{equation}
the $r_a$ are the electron coordinates,
and the sum over $a$ runs over the 
atomic electrons.
The spherical harmonic is $Y_{\ell m}$,
and the shorthand notation $\hat{r}_a$
summarizes the polar and azimuth angles $\theta_a$ and
$\varphi_a$ which define $\hat r_a$ uniquely, and vice versa.
Here, 
\begin{equation}
\left. T^{ij} \right|_{\ell = 2} =
\frac12 \, \left( T^{ij} + T^{ji} \right) -
\frac13 \, \delta^{ij} \, \trc(T) 
\end{equation}
is the traceless quadrupole component of a general 
tensor $T^{ij}$.  The octupole energy shift ($\ell = 3$) is
evaluated according to Eq.~\eqref{EEL3},
\begin{equation}
\mathcal{E}_3 = -\frac{1}{180}
\sum_{\vec{k}\lambda}\sum_{ijk}
\alpha_3(\omega)\,
\big\vert 
\left. \{ \nabla^j \nabla^k E^i_{\vec{k} \lambda}(\vec{r}) \}
\right|_{\ell = 3}  \big\vert^2 \,.
\end{equation}
The ($\ell = 3$)-component can be extracted from 
a third-rank tensor $T^{ijk}$ as follows.
One first decomposes $T^{ijk}$ into a 
totally symmetric part $S^{ijk}$, a
totally skew-symmetric part $A^{ijk}$,
and a remainder term $N^{ijk}$,
\begin{subequations}
\begin{align}
S^{ijk} =& \; T^{(ijk)} 
\\[0.1133ex]
=& \; \frac{1}{6} \left( 
T^{ijk} + T^{jki} + T^{kij} +
T^{jik} + T^{kji} + T^{ikj} \right) \,,
\nonumber\\[0.1133ex]
A^{ijk} =& \; T^{[ijk]} 
\\[0.1133ex]
=& \; \frac{1}{6} \left( 
T^{ijk} + T^{jki} + T^{kij} -
T^{jik} - T^{kji} - T^{ikj} \right) \,,
\nonumber\\[0.1133ex]
N^{ijk} =& \; T^{[ijk]} - S^{ijk} - A^{ijk} \,.
\end{align}
\end{subequations}
The matrix $S^{ijk}$ contains both 
components with $\ell = 1$ and $\ell = 3$.
One then forms a vector $\alpha^k$ which ones 
promotes to a matrix $K^{ijk}$,
\begin{align}
\alpha^k =& \; \delta^{ij} \, S^{ijk} =
\frac13 \, \left( 
\delta^{ij} \, T^{ijk} +
\delta^{ij} \, T^{ikj} +
\delta^{ij} \, T^{kij} \right) \,,
\nonumber\\[0.1133ex]
K^{ijk} =& \; 
\frac15 \, 
\left( \alpha^i \, \delta^{jk} +
\alpha^j \, \delta^{ik} +
\alpha^k \, \delta^{ij} \right) \,.
\end{align}
Finally,
the ($\ell = 3$)-component which we need for 
our considerations is obtained as
\begin{equation}
\left. T^{ijk} \right|_{\ell = 3} =
S^{ijk} - K^{ijk} \,.
\end{equation}

For the hexadecupole energy shift,
we need the ($\ell = 4$)-component of the 
Cartesian tensor
\begin{equation}
T^{ijk\ell} = 
\nabla^i \nabla^j \nabla^k E^\ell_{\vec{k} \lambda}(\vec{r}) \,.
\end{equation}
It can be calculated as follows~\cite{AnGh1982},
\begin{multline}
\label{T4}
\left. T^{ijk\ell} \right|_{\ell = 4} =
T^{(ijk\ell)}  
- \frac{1}{7} \left( 
\delta^{ij} T^{(\rho\rho k \ell)}
+ \delta^{ik} T^{(\rho\rho j \ell)}
\right.
\\
\left.
+ \delta^{i\ell} T^{(\rho\rho j k)}
+ \delta^{jk} T^{(\rho\rho i \ell)}
+ \delta^{j\ell} T^{(\rho\rho i k)}
+ \delta^{k\ell} T^{(\rho\rho i j)}
\right)
\\
+ \frac{1}{35}
\left( 
\delta^{ij} \delta^{k\ell} + 
\delta^{ik} \delta^{j\ell} +
\delta^{i\ell} \delta^{jk}
\right) T^{(\rho\rho\sigma\sigma)} \,.
\end{multline}
Here, the Einstein summation is understood,
with dummy indices ($\rho$ and $\sigma$) being
summed over. Furthermore, the 
symmetrization of a tensor is defined as follows,
\begin{multline}
T^{(ijk\ell)} =
\frac{1}{24} \,
\bigl[ 
T^{i j k \ell} 
+ T^{i j \ell k} 
+ T^{i k j \ell} 
+ T^{i k \ell j} 
+ T^{i \ell j k} 
\\
+ T^{i \ell k j} 
+ T^{j i k \ell} 
+ T^{j i \ell k} 
+ T^{j k i \ell} 
+ T^{j k \ell i} 
+ T^{j \ell i k} 
+ T^{j \ell k i} 
\\
+ T^{k i j \ell} 
+ T^{k i \ell j} 
+ T^{k j i \ell} 
+ T^{k j \ell i} 
+ T^{k \ell i j} 
+ T^{k \ell j i} 
+ T^{\ell i j k} 
\\
+ T^{\ell i k j} 
+ T^{\ell j i k} 
+ T^{\ell j k i} 
+ T^{\ell k i j} 
+ T^{\ell k j i}
\bigr]
\end{multline}
For the case $\ell = 4$, just as is the case for $\ell = 3$, 
the isolation of the 
component of highest angular momentum is a 
nontrivial exercise~\cite{AnGh1982}.

%
% Perfect Conductor
%
\subsection{Perfect Conductor}
\label{sec2B}

Based on the formalism outlined in Secs.~\ref{sec1}
and~\eqref{sec2}, and in Chap.~5 of Ref.~\cite{JeAd2022book},
it is relatively straightforward
to evaluate the atom-wall interaction
for a perfect conductor.
We first recall the known result for 
the atomic-dipole contribution,
e.g., from Eq.~(27) of Ref.~\cite{LaDKJe2010pra},
\begin{multline}
\mathcal{E}_1(z) = 
-\frac{\hbar}{16 \pi^2 \epsilon_0 z^3}
\int\limits_0^\infty \!{\rm d}\omega \,
\alpha_1(\ii\omega)
\\[0.1133ex]
\times \left[1 \!+\! \frac{2 \omega z}{c}  +
2 \left(\frac{\omega z}{c}\right)^2 \right] \!
\exp\left( -\frac{2\omega z}{c} \right) \,.
\end{multline}
The $(\ell = 2)$-contribution to the 
energy shift is as follows,
\begin{multline}
\label{quadanypol}
\mathcal{E}_2(z) =
-\frac{\hbar}{16\pi^2 \epsilon_0 z^5}
\int_0^\infty \dd\omega\,
\alpha_2(\ii \, \omega)\,
\exp\left( -\frac{2\omega z}{c} \right) 
\\[0.1133ex]
\times \left[
\frac{3}{4} +
\frac{3}{2} \, \frac{\omega z}{c} + 
\frac{4}{3} \, \left(\frac{\omega z}{c}\right)^2 +
\frac{2}{3} \, \left(\frac{\omega z}{c}\right)^3 +
\frac{1}{6} \left(\frac{\omega z}{c}\right)^4
\right] \,.
\end{multline}
This result constitutes a 
correction to a result previous 
derived in Eq.~(32) of Ref.~\cite{LaDKJe2010pra}.
For the octupole term, we have
\begin{multline}
\label{octoanypol}
\mathcal{E}_3(z) =
-\frac{\hbar}{16\pi^2 \epsilon_0 z^7}
\int_0^\infty \dd\omega\,
\alpha_3(\ii \, \omega) 
\ee^{-2\omega z/c} \,
\\[0.1133ex]
\times \left[
\frac{2}{3} +
\frac{4}{3} \, \frac{\omega z}{c} 
+ \frac{92}{75} \, \left(\frac{\omega z}{c}\right)^2 
+ \frac{152}{225} \, \left(\frac{\omega z}{c}\right)^3 
\right.
\\[0.1133ex]
\left.
+ \frac{6}{25} \, \left(\frac{\omega z}{c}\right)^3 +
\frac{4}{75} \left(\frac{\omega z}{c}\right)^4 +
\frac{4}{675} \left(\frac{\omega z}{c}\right)^6 \right] \,.
\end{multline}
For large distances, the multipole polarizabilities 
are suppressed by higher powers of $z$.
Specifically, the $2^\ell$-pole contribution
scales as $z^{-2 \ell + 1}$ in the short-range limit
and as $z^{-2 \ell + 2}$ for $z \to \infty$.
The general structure of the 
short-range asymptotic limit is given 
in Eq.~\eqref{genshort},
which in the limit $\epsilon(\ii \omega) \to \infty$
has the same functional dependence
but a different prefactor than
Eq.~(49) of Ref.~\cite{LaDKJe2010pra}.
In the context of the current investigations,
it is useful to clarify that we understand by 
short range (see also Ref.~\cite{LaDKJe2010pra})
the distance regime $a_0 \ll z \ll a_0/\alpha$,
where $a_0$ is the Bohr radius and $\alpha$ is the 
fine-structure constant. As explained in 
Ref.~\cite{ZaKo1976}, the first inequality 
$a_0 \ll z$ should
be taken with a grain of salt in this context;
the short-range expressions
derived below are actually valid down to 
distance regions of a few angstroms away 
from the surface~\cite{ZaKo1976,TaRa2014}.
From the point of view of physisorption,
what we refer to as the short-range regime 
rather constitutes a long-range distance~\cite{ZaKo1976,TaRa2014}.
By contrast, the long-range regime as considered in the 
current investigation 
refers to atom-surface distances $z \gg a_0/\alpha$.

Finally, on the basis of Eqs.~\eqref{EEL4} and~\eqref{T4},
while employing, otherwise, the formalism of Ref.~\cite{LaDKJe2010pra},
the hexadecupole energy shift
is obtained as follows,
\begin{multline}
\label{hexaanypol}
\mathcal{E}_4(z) =
-\frac{\hbar}{16\pi^2 \epsilon_0 z^9}
\int_0^\infty \dd\omega\,
\alpha_4(\ii \, \omega)
\ee^{-2\omega z/c} \,
\\[0.1133ex]
\left[
\frac{5}{8} +
\frac{5}{4} \, \frac{\omega z}{c}
+ \frac{115}{98} \, \left(\frac{\omega z}{c}\right)^2
+ \frac{100}{147} \, \left(\frac{\omega z}{c}\right)^3
\right.
\\[0.1133ex]
+ \frac{79}{294} \, \left(\frac{\omega z}{c}\right)^4
+ \frac{11}{147} \left(\frac{\omega z}{c}\right)^5 
+ \frac{32}{2205} \left(\frac{\omega z}{c}\right)^6
\\[0.1133ex]
\left.
+ \frac{4}{2205} \left(\frac{\omega z}{c}\right)^7 
+ \frac{1}{8820} \left(\frac{\omega z}{c}\right)^8
 \right] \,.
\end{multline}
This concludes the discussion of interactions
with a perfect conductor.

%
% Interactions with a Dielectric Surface
%
\section{Dielectric Surface}
\label{sec3}

%
% Dipole Term
%
\subsection{Dipole Term}
\label{sec3A}

In order to analyze the multipole contributions 
for a dielectric, one consults the 
transverse electric (TE) 
and transverse magnetic (TM) modes
given in Eqs.~(18),~(19) and (20) of Ref.~\cite{LaDKJe2010pra}.
In this case,  TE stands for incident waves whose
electric field is transverse to the plane of incidence, whereas TM 
stands for waves whose magnetic field is
transverse to the plane of incidence.
The calculation is described in detail in Ref.~\cite{LaDKJe2010pra}.
For given wave vector $\vec k$, one has two polarization 
vectors, one for the the TE mode, and another one for the TM mode.
With these ideas in mind, it is relatively easy to 
rederive the following result
for a dipole polarizable 
particle in contact with a dielectric surface,
\begin{align}
\label{EEL1res}
\mathcal{E}_1(z) =& \;
- \frac{\hbar}{8 \pi^2 \epsilon_0 \, c^3}  \, 
\int_0^\infty\dd\omega\,\omega^3\,
\alpha_1(\ii \omega) \, 
\nonumber\\[0.1133ex]
& \; \times \int_1^\infty\dd p  \,
\exp\left( - \frac{2 \,  p  \, \omega \, z}{c} \right)\, 
H( \epsilon( \ii \omega), p )\, ,
\end{align}
where
\begin{equation}
\label{defH}
H(\epsilon, p)= \frac{\sqrt{\epsilon-1+p^2}-p}
{\sqrt{\epsilon-1+p^2}+p}
+(1-2p^2)
\frac{\sqrt{\epsilon-1+p^2}-p \, \epsilon}
{\sqrt{\epsilon-1+p^2}+p \, \epsilon}.
\end{equation}
For the convenience of the reader,
a remark might be in order. Namely, in order to 
obtain Eq.~\eqref{EEL1res} from Ref.~\cite{Li1955},
one sets, in Ref.~\cite{Li1955}, $\epsilon_2(\omega) = 1 +
N_V \, \alpha_1(\omega)/\epsilon_0$, for material number 2, one expands to
first order in the volume density $N_V = N/V$, where $N$ is the number of atoms
and $V$ is the volume, and one applies the principle of virtual work. 
This leads to the potential given in Eq.~\eqref{EEL1res}.
The connection of Ref.~\cite{Li1955} 
to the atom-surface interaction was also 
pointed out in Ref.~\cite{ZaKo1976},
in Ref.~\cite{DzLiPi1961advphys},
and in the second paragraph of p.~6 of Ref.~\cite{BoKlMoMo2009}.
Specifically, after the pioneering paper~\cite{Li1955},
steps toward the calculation of the long-range and short-range limits of 
Eq.~\eqref{EEL1res} were considered
in Eqs.~(4.37)---(4.39) of Ref.~\cite{DzLiPi1961advphys},
and in Eqs.~(3) and~(4) of Ref.~\cite{DeDzKoPi1965}.
The interpolating formula~\eqref{EEL1res}
has been been given in Eqs.~(18) and (21) of Ref.~\cite{AnPiSt2004},
and in Eqs.~(63a) and (63b) of Ref.~\cite{LaDKJe2010pra}.

It is useful to investigate the function
\begin{equation}
K(\epsilon, z) = \int_1^\infty\dd p  \,
\exp\left( - \frac{2 \,  p  \, \omega \, z}{c} \right)\, 
H( \epsilon, p )\,.
\end{equation}
For $\epsilon \to \infty$, we have
$H( \epsilon, p ) \approx 2 \,  p^2$,
and 
\begin{equation}
K(\epsilon, z) \mathop{\approx}^{\epsilon \to \infty}
\frac12 \, \left( \frac{c}{z \omega}\right)^3 \,
\left[ 1 + 2 \, \frac{z \omega}{c} + 
2 \, \left( \frac{z \omega}{c} \right)^2 \right] \,,
\end{equation} 
which shows that the formula~\eqref{EEL2res} 
is consistent with the result 
for the dipole interaction with a perfect conductor.
For $z \to 0$, we have, on the other hand,
\begin{equation}
\label{Klim1}
K(\epsilon, z) =
\frac12 \, \left( \frac{c}{z \, \omega} \right)^3 \, 
\frac{\epsilon - 1}{\epsilon + 1} +
{\mathcal O}(z^{-1}) \,.
\end{equation}
Inserting \eqref{Klim1} into \eqref{EEL2res}, we have
the short-range limit as
\begin{equation}
\label{EEL1short}
\mathcal{E}_1(z) \mathop{ = }^{z \to 0}
-\frac{\hbar}{16\pi^2 \, \epsilon_0 \,z^3} 
\int_0^\infty\dd\omega\, \alpha_1(\ii \omega) \, 
\frac{\epsilon(\ii \omega)-1}{\epsilon(\ii \omega)+1}\,.
\end{equation}
The long-range limit is obtained as a 
function of the static polarizability 
$\alpha(0)$ and the static dielectric function $\epsilon(0)$
as follows,
\begin{equation}
\label{EEL1long}
\mathcal{E}_1(z) \mathop{ = }^{z \to \infty}
-\frac{3 c \hbar \, \alpha_1(0)}{32 \pi^2 \, \epsilon_0 \,z^4} \, 
\Psi_1(\epsilon(0)) \,.
\end{equation}
Here, $\Psi_1(\epsilon)$ is a function
which is normalized to unity in the limit 
$\epsilon(0) \to \infty$ (limit of a perfect
conductor) and which can otherwise be expressed
as follows,
\begin{multline}
\label{Psi1}
\Psi_1(\epsilon) = 
A_1(\epsilon) + 
B_1(\epsilon) \, 
\ln\left( \frac{ \sqrt{ \epsilon - 1 } - \sqrt{\epsilon} + 1 }%
{ \sqrt{ \epsilon - 1 } + \sqrt{\epsilon} - 1 } \right)
\\
+ C_1(\epsilon) \, 
\ln\left( \frac{ \sqrt{ \epsilon + 1 } - \sqrt{\epsilon} + 1 }%
{ \sqrt{ \epsilon + 1 } + \sqrt{\epsilon} - 1 } \right) \,.
\end{multline}
The coefficients involve both fractional and integer 
powers of $\epsilon$,
\begin{subequations}
\begin{align}
A_1(\epsilon) =& \;
\frac{ 6 \epsilon^2 - 3 \epsilon^{3/2} - 4 \epsilon
- 3 \sqrt{\epsilon} + 10 }{ 6 (\epsilon - 1)} \,,
\\[0.1133ex]
B_1(\epsilon) =& \;
\frac{ 2 \epsilon^3 - 4 \epsilon^2 + 3 \epsilon + 1 }%
{ 2 (\epsilon - 1)^{3/2} } \,,
\\[0.1133ex]
C_1(\epsilon) =& \;
- \frac{ \epsilon^2 }{ \sqrt{ \epsilon + 1} } \,.
\end{align}
\end{subequations}
The expansion about the perfect-conductor limit is
\begin{equation}
\Psi_1(\epsilon) = 
1 - \frac{5}{4 \sqrt{\epsilon}} 
+ \frac{22}{15 \epsilon} 
+ \calO\left(\frac{1}{\epsilon^{3/2}}\right) \,,
\end{equation}
which is tantamount to an expansion about a
large static dielectric function $\epsilon \equiv
\epsilon(0)$.
We reemphasize that the long-range limit is consistent with 
Eqs.~(4.37)---(4.39) of Ref.~\cite{DzLiPi1961advphys},
with Eq.~(23) of Ref.~\cite{AnPiSt2004},
and with Eqs.~(27)---(29) of Ref.~\cite{MoEtAl2022}.

%
% Quadrupole Term
%
\subsection{Quadrupole Term}
\label{sec3B}
 
The generalization to the quadrupole polarizability 
reads as follows,
\begin{multline}
\label{EEL2res}
\mathcal{E}_2(z) =
-\frac{\hbar}{16\pi^2\,\epsilon_0 \, c^5} 
\int_0^\infty\dd\omega\,\omega^5\,
\alpha_2(\ii \omega)
\\
\times 
\int_1^\infty\dd p \, \ee^{-2 p \omega z/c} 
\left(\frac{ p ^2}{2} - \frac13 \right) \, H( \epsilon(\ii \omega), p )\,.
\end{multline}
In the perfect conductor limit ($\epsilon\to\infty$),
this result is in agreement with 
the previously derived result given in Eq.~(\ref{quadanypol}). 
The short-range limit reads as follows,
\begin{equation}
\label{EEL2short}
\mathcal{E}_2(z) \mathop{ = }^{z \to 0}
-\frac{3 \hbar}{64 \pi^2 \, \epsilon_0 \,z^5}
\int_0^\infty\dd\omega\, \alpha_2(\ii \omega) \,
\frac{\epsilon(\ii \omega)-1}{\epsilon(\ii \omega)+1}\,.
\end{equation}
The long-range limit is obtained as follows,
\begin{equation}
\label{EEL2long}
\mathcal{E}_2(z) \mathop{ = }^{z \to \infty}
-\frac{35 c \hbar \, \alpha_2(0)}{384 \pi^2 \, \epsilon_0 \,z^6} \, 
Psi_2(\epsilon(0)) \,,
\end{equation}
where $\Psi_2$ has the same structure as $\Psi_1$,
\begin{multline}
\label{Psi2}
\Psi_2(\epsilon) = 
A_2(\epsilon) + 
B_2(\epsilon) \, 
\ln\left( \frac{ \sqrt{ \epsilon - 1 } - \sqrt{\epsilon} + 1 }%
{ \sqrt{ \epsilon - 1 } + \sqrt{\epsilon} - 1 } \right)
\\
+ C_2(\epsilon) \, 
\ln\left( \frac{ \sqrt{ \epsilon + 1 } - \sqrt{\epsilon} + 1 }%
{ \sqrt{ \epsilon + 1 } + \sqrt{\epsilon} - 1 } \right) \,.
\end{multline}
The coefficients are given as follows,
\begin{subequations}
\begin{align}
A_2(\epsilon) & =
\frac{1}{ 140 (\epsilon - 1)^2 } 
\left[ -120 \epsilon^4 + 60 \epsilon^{7/2} 
+ 380 \epsilon^3 
\right.
\nonumber\\[0.1133ex]
& \left. 
- 180 \epsilon^{5/2} 
- 364 \epsilon^2 + 75  \epsilon^{3/2} + 348 \epsilon
+ 75 \sqrt{\epsilon} - 224 \right] \,,
\\[0.1133ex]
B_2(\epsilon) & =
\frac{3}{28} 
\frac{ 8 \epsilon^5 - 28 \epsilon^4 + 40 \epsilon^3 
- 34 \, \epsilon^2 + 7 \epsilon + 5 }%
{ (\epsilon - 1)^{5/2} } \,,
\\[0.1133ex]
C_2(\epsilon) & =
\frac37 \, \frac{ \epsilon^2 ( 2 \epsilon - 1 )}{ \sqrt{ \epsilon + 1 } } \,.
\end{align}
\end{subequations}
The expansion about the perfect-conductor limit is
\begin{equation}
\Psi_2(\epsilon) = 
1 - \frac{31}{28 \sqrt{\epsilon}} 
+ \frac{338}{245 \epsilon} 
+ \calO\left(\frac{1}{\epsilon^{3/2}}\right) \,.
\end{equation}

%
% Octupole Term
%
\subsection{Octupole Term}
\label{sec3C}

For the interaction with a dielectric surface,
the octupole energy shift reads as follows,
\begin{multline}
\label{EEL3res}
\mathcal{E}_3(z) =
-\frac{\hbar}{16 \pi^2 \, \epsilon_0 \,c^7} 
\int_0^\infty\dd\omega\,\omega^7
\alpha_3(\ii \omega)
\int_1^\infty\dd p  
\ee^{-2 p \omega z/c} 
\\
\times \left( \frac{8}{135} \, p^4
- \frac{16}{225} \, p^2 + \frac{4}{225} \right) \,
H( \epsilon(\ii \omega), p) \,.
\end{multline}
The calculation of the dielectric response function for 
imaginary input frequencies is a nontrivial problem.
The short-range limit is given by the expression
\begin{equation}
\label{EEL3short}
\mathcal{E}_3(z) \mathop{ = }^{z \to 0}
-\frac{\hbar}{24 \pi^2 \, \epsilon_0 \,z^7}
\int_0^\infty\dd\omega\, \alpha_3(\ii \omega) \,
\frac{\epsilon(\ii \omega)-1}{\epsilon(\ii \omega)+1}\,.
\end{equation}
The long-range limit is 
\begin{equation}
\label{EEL3long}
\mathcal{E}_3(z) \mathop{ = }^{z \to \infty}
-\frac{77 c \hbar \, \alpha_3(0)}{800 \pi^2 \, \epsilon_0 \,z^8} \, 
\Psi_3(\epsilon(0)) \,,
\end{equation}
where $\Psi_3$ reads as follows,
\begin{multline}
\label{Psi3}
\Psi_3(\epsilon) = 
A_3(\epsilon) + 
B_3(\epsilon) \, 
\ln\left( \frac{ \sqrt{ \epsilon - 1 } - \sqrt{\epsilon} + 1 }%
{ \sqrt{ \epsilon - 1 } + \sqrt{\epsilon} - 1 } \right)
\\
+ C_3(\epsilon) \, 
\ln\left( \frac{ \sqrt{ \epsilon + 1 } - \sqrt{\epsilon} + 1 }%
{ \sqrt{ \epsilon + 1 } + \sqrt{\epsilon} - 1 } \right) \,.
\end{multline}
The coefficients are given as follows,
\begin{subequations}
\begin{align}
A_3(\epsilon) & =
\frac{1}{ 11088 (\epsilon - 1)^3 } 
\biggl[ 
-2520 \epsilon^{11/2} 
-28560 \epsilon^5 
\nonumber\\[0.1133ex]
& \; 
+ 13860 \epsilon^{9/2} 
+ 66528 \epsilon^4 
- 31290 \epsilon^{7/2} 
\nonumber\\[0.1133ex]
& \; 
- 76336 \epsilon^3 
+ 29120 \epsilon^{5/2} 
+ 62016 \epsilon^2 
\nonumber\\[0.1133ex]
& \; 
- 2415 \epsilon^{3/2} 
- 48675 \epsilon 
- 6195 \sqrt{\epsilon} + 19888 \biggr] \,,
\\[0.1133ex]
B_3(\epsilon) & =
\frac{5}{528 (\epsilon - 1)^{7/2} } 
\bigl[ 
48 \epsilon^7 - 288 \epsilon^6
+ 712 \epsilon^5 
\nonumber\\[0.1133ex]
& \; - 976 \epsilon^4
+ 890 \epsilon^3 - 444 \epsilon^2
+ 5 \epsilon + 59 \bigr] \,,
\\[0.1133ex]
C_3(\epsilon) & = - \frac{5}{33} \, 
\frac{ 3 \epsilon^2 - 6 \epsilon + 1 }{ \sqrt{ \epsilon + 1 } } \,.
\end{align}
\end{subequations}
The expansion about the perfect-conductor limit is
\begin{equation}
\Psi_3(\epsilon) = 
1 - \frac{265}{264 \sqrt{\epsilon}} 
+ \frac{914}{693 \epsilon} 
+ \calO\left(\frac{1}{\epsilon^{3/2}}\right) \,.
\end{equation}

%
% Hexadecupole Term
%
\subsection{Hexadecupole Term}
\label{sec3D}
 
For the interaction with a dielectric surface,
the octupole energy shift reads as follows,
\begin{multline}
\label{EEL4res}
\mathcal{E}_4(z) =
-\frac{\hbar}{16 \pi^2 \, \epsilon_0 \,c^9}
\int_0^\infty\dd\omega\,\omega^9
\alpha_4(\ii \omega)
\int_1^\infty\dd p
\ee^{-2 p \omega z/c}
\\
\times \left( 
\frac{1}{252} \, p^6
- \frac{1}{147} \, p^4
+ \frac{1}{294} \, p^2 - \frac{1}{2205} \right) \,
H( \epsilon(\ii \omega), p) \,.
\end{multline}
The short-range limit is 
\begin{equation}
\label{EEL4short}
\mathcal{E}_4(z) \mathop{ = }^{z \to 0}
-\frac{5 \hbar}{128 \pi^2 \, \epsilon_0 \,z^9}
\int_0^\infty\dd\omega\, \alpha_4(\ii \omega) \,
\frac{\epsilon(\ii \omega)-1}{\epsilon(\ii \omega)+1}\,.
\end{equation}
The long-range limit is obtained as follows,
\begin{equation}
\label{EEL4long}
\mathcal{E}_4(z) \mathop{ = }^{z \to \infty}
-\frac{1287 c \hbar \, \alpha_4(0)}{12544 \pi^2 \, \epsilon_0 \,z^{10}} \, 
\Psi_4(\epsilon(0)) \,,
\end{equation}
where $\Psi_4$ reads as follows,
\begin{multline}
\label{Psi4}
\Psi_4(\epsilon) = 
A_4(\epsilon) + 
B_4(\epsilon) \, 
\ln\left( \frac{ \sqrt{ \epsilon - 1 } - \sqrt{\epsilon} + 1 }%
{ \sqrt{ \epsilon - 1 } + \sqrt{\epsilon} - 1 } \right)
\\
+ C_4(\epsilon) \, 
\ln\left( \frac{ \sqrt{ \epsilon + 1 } - \sqrt{\epsilon} + 1 }%
{ \sqrt{ \epsilon + 1 } + \sqrt{\epsilon} - 1 } \right) \,.
\end{multline}
The coefficients are given as follows,
\begin{subequations}
\begin{align}
A_4(\epsilon) & =
\frac{1}{ 205920 (\epsilon - 1)^4 } 
\biggl[ 
- 40320 \epsilon^{8} 
+ 20160 \epsilon^{15/2}
\nonumber\\[0.1133ex]
& \; 
+ 369600 \epsilon^{7} 
- 181440 \epsilon^{13/2}
- 1338624 \epsilon^{6} 
\nonumber\\[0.1133ex]
& \; 
+ 641760 \epsilon^{11/2}
+ 2670816 \epsilon^{5} 
- 1252440 \epsilon^{9/2} 
\nonumber\\[0.1133ex]
& \; 
- 3242464 \epsilon^4 
+ 1409520 \epsilon^{7/2}
+ 2767936 \epsilon^{3}
\nonumber\\[0.1133ex]
& \;
- 690270 \epsilon^{5/2} 
- 2110464 \epsilon^{2}
- 62265 \epsilon^{3/2}
\nonumber\\[0.1133ex]
& \;
+ 1300000 \epsilon
+ 118125 \epsilon^{1/2}
- 376480 \biggr] \,,
\\[0.1133ex]
B_4(\epsilon) & =
-\frac{7}{4576 (\epsilon - 1)^{9/2} } 
\bigl[ 
128 \epsilon^9 - 1216 \epsilon^8
+ 4608 \epsilon^7
\nonumber\\[0.1133ex]
& \; - 9600 \epsilon^6
+ 12720 \epsilon^5 - 12120 \epsilon^4
\nonumber\\[0.1133ex]
& \; + 8080 \epsilon^3
- 2580 \epsilon^2 - 405 \epsilon
+ 375 \bigr] \,,
\\[0.1133ex]
C_4(\epsilon) & = \frac{7}{143} \, 
\frac{ 4 \epsilon^3 - 18 \epsilon^2 + 12 \epsilon - 1}{ \sqrt{ \epsilon + 1 } } \,.
\end{align}
\end{subequations}
The expansion about the perfect-conductor limit is
\begin{equation}
\Psi_4(\epsilon) = 
1 - \frac{1323}{1430 \sqrt{\epsilon}} 
+ \frac{18050}{14157 \epsilon} 
+ \calO\left(\frac{1}{\epsilon^{3/2}}\right) \,.
\end{equation}

%
% General Short--Range Asymptotics
%
\subsection{Some General Short--Range Results}
\label{sec3E}

In the short-range limit,
a few simplifications and generalizations
are possible, especially in regard to 
two-body bound systems like hydrogen and 
positronium.
First, we may point out the 
generalization of the short-range
expressions given in 
Eqs.~\eqref{EEL1short},~\eqref{EEL2short},~\eqref{EEL3short},
and~\eqref{EEL4short}, to arbitrary multipole orders.
Indeed, the general result for the $2^\ell$-pole effect
reads as follows,
\begin{multline}
\label{genshort}
\mathcal{E}_\ell(z) \mathop{ = }^{z \to 0}
- \frac{\hbar}{16 \pi^2 \, \epsilon_0 \,z^{2 \ell +1}} \,
\\
\times 
\frac{\ell + 1}{2 \ell} 
\int_0^\infty\dd\omega\, \alpha_\ell(\ii \omega) \,
\frac{\epsilon(\ii \omega)-1}{\epsilon(\ii \omega)+1}\,.
\end{multline}
This result has the same functional form as 
Eq.~(49) of Ref.~\cite{LaDKJe2010pra},
but a different, updated prefactor.

\begin{figure*}[t!]
\begin{center}
\begin{minipage}{1.0\linewidth}
\begin{center}
\includegraphics[width=0.91\linewidth]{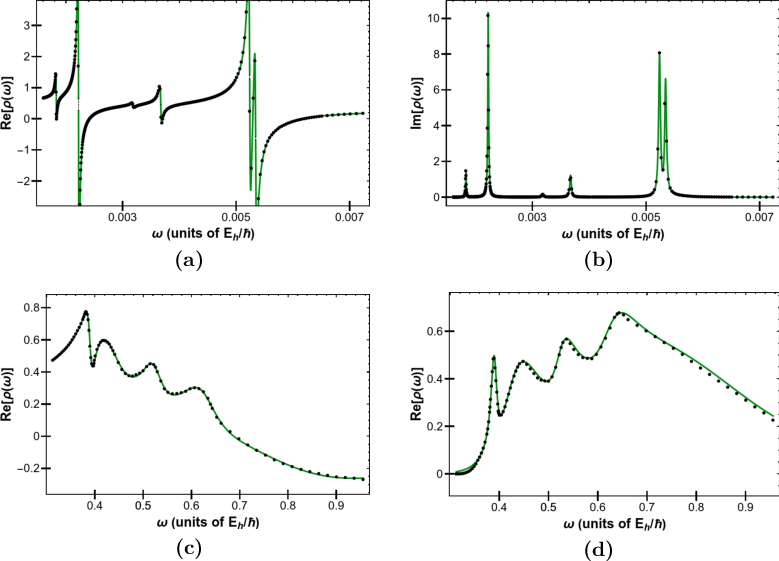}
\caption{\label{fig1}
A reanalysis of the dielectric function
of $\alpha$-quartz (ordinary axis) 
is performed based on Eq.~\eqref{new_fit_formula}, with fitting parameters 
given in Table~\ref{table1}.
The angular frequency is measured in 
atomic units, i.e., in units of $E_h/\hbar$,
where $E_h$ is the Hartree energy and $\hbar$ is the 
reduced Planck constant. The panels refer to 
(a) lattice resonance region, real part of $\rho(\omega)$,
(b) lattice resonance region, imaginary part of $\rho(\omega)$,
(c) interband region, real part of $\rho(\omega)$,
(d) interband region, imaginary part of $\rho(\omega)$.
The data points are taken from Ref.~\cite{Ph1985},
while the solid curve is described 
by Eq.~\eqref{new_fit_formula}, with parameters given in Table~\ref{table1}.}
\end{center}
\end{minipage}
\end{center}
\end{figure*}

\begin{figure*}[t!]
\begin{center}
\begin{minipage}{1.0\linewidth}
\begin{center}
\includegraphics[width=0.91\linewidth]{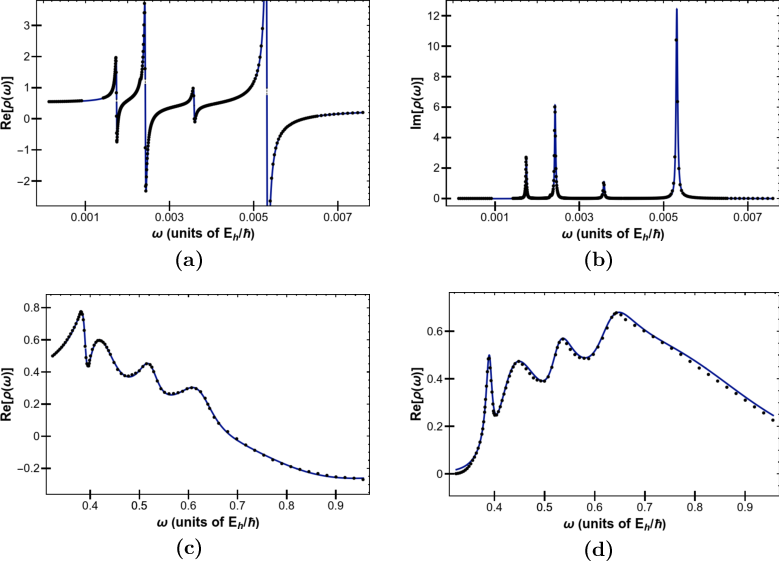}
\caption{\label{fig2}
We present the analogue of Fig.~\ref{fig1}, but for the 
extraordinary axis. Specifically, the dielectric function
of $\alpha$-quartz (extraordinary axis)
is analyzed based on Eq.~\eqref{new_fit_formula}, with fitting parameters
given in Table~\ref{table1}.
The angular frequency is measured in 
atomic units, i.e., in units of $E_h/\hbar$,
where $E_h$ is the Hartree energy and $\hbar$ is the
reduced Planck constant. The panels refer to
(a) lattice resonance region, real part of $\rho(\omega)$,
(b) lattice resonance region, imaginary part of $\rho(\omega)$,
(c) interband region, real part of $\rho(\omega)$,
(d) interband region, imaginary part of $\rho(\omega)$.
The data points are taken from Ref.~\cite{Ph1985},
while the solid curve is described 
by Eq.~\eqref{new_fit_formula}, with parameters given in Table~\ref{table1}.}
\end{center}
\end{minipage}
\end{center}
\end{figure*}

\def\arraystretch{1.25}
\begin{table}[t!]
\caption{\label{table1}
We indicate the coefficients for the first few resonances 
for $\alpha$-quartz according to 
the fitting formula~\eqref{new_fit_formula},
for the ordinary and the extraordinary optical axis. 
The values for $\omega_k$, $\gamma_k$ and $\gamma'_k$
are measured in atomic units, i.e., in units of $E_h/\hbar$.
The subscript $k$ numbers the resonances.}
\begin{center}
\begin{tabular}{c D{,}{.}{10} D{,}{.}{10} D{,}{.}{10} D{,}{.}{10} }
\hline
\hline
\multicolumn{5}{c}{Ordinary Axis: Vibrational Excitations} \\
 $k$
 & \multicolumn{1}{c}{$\alpha_k$}
 & \multicolumn{1}{c}{$\omega_k$}
 & \multicolumn{1}{c}{$\gamma_k$} 
 & \multicolumn{1}{c}{$\gamma'_k$} 
\\
\hline
 1  & 1,045 \times 10^{-2} & 1,827 \times 10^{-3} & 1,301 \times 10^{-5} &  1,631 \times 10^{-5} \\
 2  & 8,541 \times 10^{-2} & 2,218 \times 10^{-3} & 1,835 \times 10^{-5} & -1,528 \times 10^{-6} \\
 3  & 2,022 \times 10^{-3} & 3,180 \times 10^{-3} & 4,026 \times 10^{-5} &  1,019 \times 10^{-4} \\
 4  & 1,111 \times 10^{-2} & 3,668 \times 10^{-3} & 3,381 \times 10^{-5} &  3,423 \times 10^{-5} \\
 5  & 5,850 \times 10^{-2} & 5,234 \times 10^{-3} & 3,959 \times 10^{-5} & -3,881 \times 10^{-5} \\
 6  & 4,472 \times 10^{-2} & 5,339 \times 10^{-3} & 3,776 \times 10^{-5} &  4,041 \times 10^{-4} \\
\hline
\multicolumn{5}{c}{Ordinary Axis: Interband Excitations} \\
 $k$
 & \multicolumn{1}{c}{$\alpha_k$}
 & \multicolumn{1}{c}{$\omega_k$}
 & \multicolumn{1}{c}{$\gamma_k$}
 & \multicolumn{1}{c}{$\gamma'_k$} \\ 
\hline
  7  & 1,341 \times 10^{-2} & 3,899 \times 10^{-1} & 1,423 \times 10^{-2} & -1,195 \times 10^{-1} \\
  8  & 8,086 \times 10^{-2} & 4,287 \times 10^{-1} & 9,996 \times 10^{-2} &  4,432 \times 10^{-1} \\
  9  & 1,568 \times 10^{-2} & 5,245 \times 10^{-1} & 5,322 \times 10^{-2} &  7,333 \times 10^{-1} \\
 10  & 2,520 \times 10^{-1} & 6,270 \times 10^{-1} & 9,280 \times 10^{-2} &  8,041 \times 10^{-1} \\
 11  & 1,772 \times 10^{-1} & 8,237 \times 10^{-1} & 4,981 \times 10^{-1} & -4,361 \times 10^{-1} \\
\hline
\hline
\multicolumn{5}{c}{Extraordinary Axis: Vibrational Excitations} \\
 $k$
 & \multicolumn{1}{c}{$\alpha_k$}
 & \multicolumn{1}{c}{$\omega_k$}
 & \multicolumn{1}{c}{$\gamma_k$}
 & \multicolumn{1}{c}{$\gamma'_k$} \\ 
\hline
  1 & 3,628 \times 10^{-2} & 1,736 \times 10^{-3} & 2,316 \times 10^{-5} &  2,316 \times 10^{-5} \\
  2 & 6,734 \times 10^{-2} & 2,418 \times 10^{-3} & 2,811 \times 10^{-5} &  4,267 \times 10^{-5} \\
  3 & 1,023 \times 10^{-2} & 2,430 \times 10^{-3} & 9,981 \times 10^{-5} & -3,154 \times 10^{-3} \\
  4 & 1,114 \times 10^{-2} & 3,578 \times 10^{-3} & 3,594 \times 10^{-5} &  2,118 \times 10^{-3} \\
  5 & 1,026 \times 10^{-1} & 5,307 \times 10^{-3} & 4,383 \times 10^{-5} &  2,533 \times 10^{-4} \\
\hline
\multicolumn{5}{c}{Extraordinary Axis: Interband Excitations} \\
 $k$
 & \multicolumn{1}{c}{$\alpha_k$}
 & \multicolumn{1}{c}{$\omega_k$}
 & \multicolumn{1}{c}{$\gamma_k$}
 & \multicolumn{1}{c}{$\gamma'_k$} \\
\hline
  6  & 1,351 \times 10^{-2} & 3,899 \times 10^{-1} & 1,428 \times 10^{-2} & -1,172 \times 10^{-1} \\
  7  & 7,775 \times 10^{-2} & 4,283 \times 10^{-1} & 9,861 \times 10^{-2} &  4,422 \times 10^{-1} \\
  8  & 1,512 \times 10^{-2} & 5,245 \times 10^{-1} & 5,229 \times 10^{-2} &  7,323 \times 10^{-1} \\
  9  & 2,430 \times 10^{-1} & 6,272 \times 10^{-1} & 9,108 \times 10^{-2} &  7,971 \times 10^{-1} \\
 10  & 1,859 \times 10^{-1} & 8,209 \times 10^{-1} & 5,049 \times 10^{-1} & -4,160 \times 10^{-1} \\
\hline
\hline
\end{tabular}
\end{center}
\end{table}

\begin{table}[t!]
\caption{\label{table2} 
Coefficient $C_{(2\ell+1)0}$ multiplying the leading term 
for the $2^\ell$-pole contribution to the 
atom-surface interaction are given
in atomic units, for hydrogen 
interacting with a (perfect) conductor,
and with $\alpha$-quartz.}
\begin{center}
\begin{tabular}{l@{\hspace{0.3cm}}r@{\hspace{0.3cm}}r@{\hspace{0.3cm}}r@{\hspace{0.3cm}}r}
\hline
\hline
\multicolumn{5}{c}{$C_{(2\ell+1)0}$ for Hydrogen} \\
& $\ell=1$ & $\ell=2$ & $\ell=3$ & $\ell=4$ \\
\hline
Conductor        & 0.250  & 0.844 & 7.50 & 123 \\
$\alpha$--Quartz & 0.0599 & 0.178 & 1.48 & 23.4 \\
\hline
\hline
\end{tabular}
\end{center}
\end{table}

\def\arraystretch{1.25}
\begin{table}[t!]
\caption{\label{table3} 
Same as Table~\ref{table2} for positronium.}
\begin{center}
\begin{tabular}{l@{\hspace{0.3cm}}r@{\hspace{0.3cm}}r@{\hspace{0.3cm}}r@{\hspace{0.3cm}}r}
\hline
\hline
\multicolumn{5}{c}{$C_{(2\ell+1)0}$ for Positronium} \\
& $\ell=1$ & $\ell=2$ & $\ell=3$ & $\ell=4$ \\
\hline
Conductor        & 1.00 & 13.5 & 480 & 31500 \\
$\alpha$--Quartz & 0.302 & 3.73 & 127 & 8130 \\
\hline
\hline
\end{tabular}
\end{center}
\end{table}

%
% Positronium and $\alpha$-quartz
%
\section{Calculation of Multipole Corrections}
\label{sec4}

%
% Positronium and $\alpha$-quartz
%
\subsection{Multipoles 
for Hydrogen and Positronium}
\label{sec4A}

We aim to give an update on 
multipole corrections to atom-surface interactions,
beyond the discussion in Ref.~\cite{LaDKJe2010pra}.
Hydrogen and positronium constitute
atomic systems for which the exact
evaluation of multipole polarizabilities
is possible analytically (see Appendix~\ref{appa}).
Hence, we focus on these two atomic systems,
for definiteness, while
stressing that other atomic systems
could be more interesting from
the point of view of applications
(see Sec.~\ref{sec4C}).
The derivation of the multipole polarizabilities 
of hydrogen and positronium uses the 
Sturmian decomposition of the Schr\"{o}dinger--Coulomb 
Green function, and the evaluation of radial
matrix elements, according to the formalism
outlined in Chap.~4 of Ref.~\cite{JeAd2022book}. 
In order to ensure concise formulas, we now switch to atomic units
(see Chap. 2 of Ref. [13]) with $a_0 = 1$, $E_h = 1$,
$\epsilon_0 = 1/(4 \pi)$, $\hbar = 1$, and $e = 1$ (unit elementary charge).

Using the results given in Appendix~\ref{appa},
we are now in the position to derive some more closed-form
expressions for interactions with a perfect
conductor, i.e., in the limit
$\epsilon(\ii \omega) \to \infty$.
Specifically, for multipole interactions with atomic hydrogen,
one obtains the following result
for the integral over the $2^\ell$-pole polarizability,
\begin{equation}
\int_0^\infty\dd\omega\, \alpha_\ell(\ii \omega) = 
\frac{\pi \, \Gamma(2 \ell + 3)}{2^{2 \ell + 1} \, (2 \ell + 1)}
\frac{a_0^{2 \ell}}{\hbar} \,.
\end{equation}
The general result for the $2^\ell$-pole
energy shift for hydrogen (H) interacting 
with a perfect conductor thus is as follows,
\begin{multline}
\label{genmultipoleH}
\mathcal{E}^{({\rm H})}_\ell(z) \mathop{ = }^{z \to 0, \epsilon \to \infty}
-\frac{(\ell + 1) \, \Gamma(2 \ell + 3)}{2^{2 \ell + 4} \, \ell \, 
(2 \ell + 1)} \, \frac{E_h}{(z/a_0)^{2 \ell +1}} \,.
\end{multline}
Here, $E_h$ is the Hartree energy,
and $a_0$ is the Bohr radius. 
The ratio $z/a_0$ is equal to the 
atom-wall distance, expressed in atomic units
(see Chap.~2 of Ref.~\cite{JeAd2022book}).
For positronium (Ps), one obtains the result
\begin{multline}
\label{genmultipolePs}
\mathcal{E}^{({\rm Ps})}_\ell(z) \mathop{ = }^{z \to 0, \epsilon \to \infty}
-\frac{(\ell + 1) \, \Gamma(2 \ell + 3)}{16 \, \, \ell \,
(2 \ell + 1)} \, \frac{E_h}{(z/a_0)^{2 \ell +1}} \,.
\end{multline}
In the short-range regime,
the expansion into multipoles
constitutes an expansion in powers of
$a_0/z$, where $a_0$ is the Bohr
radius and $z$ is the atom-wall distance.

The general results given in 
Eqs.~\eqref{genmultipoleH} 
and~\eqref{genmultipolePs} 
demonstrate that the sum over 
the multipole potentials  (at least for hydrogen 
and positronium interacting with 
a perfect conductor) 
constitutes a divergent, asymptotic series.
The divergence, for any distance $z$, 
happens due to the factorial growth of the prefactor
$\Gamma(2 \ell + 3)$.
Optimal truncation of the multipole 
expansion at the smallest term
of the series then constitutes a valid
procedure for obtaining theoretical 
predictions~\cite{CaEtAl2007},
while Borel summation can be used in order
to sum the divergent series~\cite{Je2000prd}.

%
% Dielectric Function of $\maybebm{\alpha}$-quartz
%
\subsection{Hydrogen, Positronium and $\maybebm{\alpha}$--Quartz}
\label{sec4B}

We aim to combine the analysis of the 
multipole polarizabilities given in
Appendix~\ref{appa} and
Sec.~\ref{sec4A} with an update on 
$\alpha$-quartz (Ref.~\cite{LaDKJe2010pra}).
For the data presented in~\cite{Ph1985},
we used the following fit formula 
discussed in Ref.~\cite{LaDKJe2010pra},
\begin{align}
\label{fit_formula}
\rho(\omega) \equiv & \; \frac{\epsilon(\omega)-1}{\epsilon(\omega)+2}
= \frac{[n(\omega) + \ii \,  k(\omega)]^2-1}{[n(\omega)+\ii \, k(\omega)]^2+2}
\nonumber\\[2ex]
\simeq & \; \sum_{k=1}^n \frac{\alpha_k \, \omega_k^2}%
{\omega_k^2 - \ii \, \gamma_k \, \omega-\omega^2} \,.
\end{align}
Here, $n(\omega)$ are $k(\omega)$ are 
the dispersive and absorptive parts of the 
index of refraction, 
while the functional form 
is inspired by the Clausius--Mossotti equation.
We take the opportunity to 
point out the missing factor $\omega_k^2$ in the 
numerator of the fitting function
given in Eq.~(70) of Ref.~\cite{LaDKJe2010pra}.
The missing prefactor had previously been 
supplemented in Eq.~(21) of Ref.~\cite{JeJaDK2016pra}.
For completeness, it might be 
useful to point out that
the expression $(\epsilon(\omega)-1)/(\epsilon(\omega)+1)$,
which appears in the integrand of the
short-range 
expressions~\eqref{EEL1short},~\eqref{EEL2short},~\eqref{EEL3short},
and~\eqref{EEL4short},
can be obtained from $\rho(\omega)$ as follows,
\begin{equation}
\frac{\epsilon(\omega)-1}{\epsilon(\omega)+1} =
\frac{ 3 \rho(\omega) }{ \rho(\omega) + 2 } \,.
\end{equation}
For intrinsic silicon~\cite{MoEtAl2022},
we have recently found that a better
analytic representation can be 
obtained based on the following fit formula,
\begin{align}
\label{new_fit_formula}
\rho(\omega) 
\simeq & \; \sum_{k=1}^n \frac{\alpha_k \, 
(\omega_k^2 - \ii \, \gamma'_k \, \omega)}%
{\omega_k^2 - \ii \, \gamma_k \, \omega-\omega^2} \,.
\end{align}
where the expression $[\alpha_k \,
(\omega_k^2 - \ii \, \gamma'_k \, \omega)]$
can be regarded as a complex oscillator 
strength that includes a 
phenomenological model of radiative reaction~\cite{MoEtAl2022}.
We find 
that such a model represents the data
from Ref.~\cite{Ph1985} very well
(see Figs.~\ref{fig1} and~\ref{fig2},
as well as Table~\ref{table1}).
This finding is nontrivial in view of the 
necessity for any fit function to 
fulfill the Kramers--Kronig relationships
(see, e.g., Chap.~6 of Ref.~\cite{Je2017book}),
which relate the real and imaginary 
parts of $\rho$ and $\epsilon$.
The functional form of our model~\eqref{new_fit_formula} fulfills
the Kramers--Kronig relationships automatically.
While the oscillator strengths and resonance
frequencies for vibrational
excitations differ in between the ordinary 
and the extraordinary axis of $\alpha$-quartz
(see Table~\ref{table1}), 
we find that the influence of the 
low-frequency vibrational excitations on the leading short-range
expansion coefficients reported below is
numerically negligible.
Similar approaches as described by Eq.~\eqref{new_fit_formula}
have been discussed (for rutile) in Eq.~(1) of Ref.~\cite{Ri1985},
in Eq.~(4) of Ref.~\cite{Tr1985} (for cubic thallium),
in Eq.~(1) of Ref.~\cite{PaKh1985} (for sodium nitrate),
and in Eq.~(5) of Ref.~\cite{FuDoQu1985} (for orthorhombic sulphur).

We are now in the position to analyze 
the multipole corrections for $\alpha$-quartz.
In general, we can say that 
these intriguing corrections to atom-surface interactions
have generated considerable interest
[see, e.g., Ref.~\cite{LaDKJe2011LNP}, 
Eq.~(2) of Ref.~\cite{TaRa2014},
and Eq.~(8) of Ref.~\cite{ZhTaRa2017}].
In order to put this finding into 
perspective, we should point out that it 
has recently become possible to 
calculate the multipole corrections
to polarizabilities 
more accurately from first principle, 
for general multi-electron
atoms~\cite{KaSiArSa2022}.
In this article, in order to facilitate the 
analysis of the multipole corrections, 
we have evaluated exact expressions
for hydrogen and positronium up to the 
hexadecupole order (see Appendix~\ref{appa} and Sec.~\ref{sec4A}).

It is indicated to include a brief
discussion on the magnitude of the 
multipole corrections. We continue to use 
atomic units. {\em A priori}, the short-range approximations are
valid for $z \leq 1/\alpha \approx 137.036$
in atomic units, i.e., for distances smaller than
about 137~Bohr radii~\cite{LaDKJe2010pra,JeMo2023blog}.
The upper end of the range of validity of the short-range,
nonretarded approximation and its dependence
on the atomic species has recently 
been discussed in Ref.~\cite{DaUlJe2023}.
The leading term for the $2^\ell$-pole multipole term
is proportional to $1/z^{2 \ell + 1}$.
It is well known that Lifshitz theory 
cannot be used for arbitrarily close 
approach to the surface.
The probability density of
atomic wave functions (for ground-state hydrogen atoms)
decrease with a probability
$| \psi |^2 \sim \exp(-2 r)$,
where $r$ is the distance from the
nucleus in atomic units.
For a distance of $z = 0.5 \, {\rm nm}$,
which is roughly equal to
$z = 10$ in atomic units,
one has $| \psi |^2 \sim 10^{-9}$,
eliminating the overlap as 
a possible limiting factor.
It has been stressed in Ref.~\cite{ZaKo1976} that, for vanishing 
overlap of the atomic wave function with 
the surface, the exchange of electrons between the 
atom and the substrate can 
be neglected, which, in turn, 
makes it possible to treat the electrons
(and nuclei) of the atom and the surface 
as distinguishable.
Furthermore, in the seminal paper~\cite{ZaKo1976},
it has been stressed in remarks following
Eq.~(2.39) that even for separations 
typically encountered in physisorption
($\sim$ 4-7 Bohr radii), the formula
$V(z) \approx - C_3/(z - z_0)^3$,
is applicable. 
Here, $C_3$ is given implicitly in Eq.~\eqref{EEL1short}.
Here, $z_0$ is the position
of a suitably defined reference plane 
given by Eq.~(2.38) of Ref.~\cite{ZaKo1976}.
The position of the reference plane 
(see also Fig.~1 of Ref.~\cite{ZaKo1976})
is given by an integral which depends on 
both the susceptibility of the atom 
and also [via the integral 
given in Eq.~(2.28) of Ref.~\cite{ZaKo1976}]
on the susceptibility of the solid.
One accepted path toward the calculation
of physisorption energies has involved 
the addition of an ultrashort-range
(overlap) contribution to the 
energy, which is calculated on the basis
of density-functional (DFT) theory,
and the van-der-Waals energy,
the latter being
calculated according to the ideas 
outlined in Ref.~\cite{ZaKo1976}.
One possible pathway toward the calculation
of the contact contribution (the DFT part) is based on 
DFT--GGA, where GGA stands for the generalized
gradient approximation~\cite{PeBuEr1996}.
The entire procedure is often referred to as 
the van-der-Waals-corrected DFT 
approach~\cite{Gr2004,Gr2006,Gr2010,SiAmGrAn2012}. 
A calculation of the reference-plane position $z_0$
for $\alpha$-quartz is beyond the scope of 
the current paper.
We merely use an exemplary distance of
$z = 10$ atomic units for the calculations
reported below, in order to illustrate the 
magnitude of the multipole corrections
for close approach,
and furthermore, assume positronium 
to be at rest (cf.~Refs.~\cite{MaRi1984,PaPa1985}).

We can thus use an exemplary distance of
$z = 10$ atomic units, in order to 
analyze the magnitude of the multipole
corrections.
A further remark is in order.
It has recently been shown in Ref.~\cite{JeMo2023blog}
that higher-order terms in the atom-surface
potential contain logarithms of the 
atom-wall distance, $\ln(z)$,
leading to a semi-analytic expansion
of the atom-surface potential in powers
of $z$ and $\ln(z)$, described by coefficients
with two indices.
Hence, we will refer to the leading short-range
coefficient multiplying for the atom-surface dipole 
and quadrupole terms as $C_{30}$ and $C_{50}$
(rather than $C_3$ and $C_5$),
where the first index counts the power of $z$
and the second index (equal to zero) indicates the 
absence of a logarithm~\cite{JeMo2023blog}.

For a perfect conductor, we thus write 
the relations
\begin{equation}
\calE_1(z) \approx - \frac{C_{30}}{z^3} \,, 
\quad
\calE_2(z) \approx - \frac{C_{50}}{z^5}  \,,
\quad
1 \ll z \ll \alpha^{-1} \,,
\end{equation}
with the following exemplary results 
(for the dipole and quadrupole coefficients)
for perfect conductors
[see Eq.~\eqref{genmultipoleH} and~\eqref{genmultipolePs}]
\begin{align}
C_{30}^{({\rm H})} 
\;
\mathop{=}^{\epsilon \to \infty} 
\;
& \; \frac14 \,,
\qquad
& C_{50}^{({\rm H})} 
\;
\mathop{=}^{\epsilon \to \infty} 
\;
\frac{27}{32} \,,
\\[1.133ex]
C_{30}^{({\rm Ps})} 
\;
\mathop{=}^{\epsilon \to \infty} 
\;
& \; 1 \,,
\qquad
& C_{50}^{({\rm Ps})}  
\;
\mathop{=}^{\epsilon \to \infty} 
\;
\frac{27}{2} \,.
\end{align}
For $\alpha$-quartz, we refer to Tables~\ref{table2} and~\ref{table3}
for the multipole coefficients.
For positronium, at $z = 10$ atomic units, one has the 
following ratio of the quadrupole corrections
to the leading dipole term,
\begin{align}
\label{ratioPs}
\frac{\calE^{({\rm Ps})}_2(z = 10)}{\calE^{({\rm Ps})}_1(z = 10)}
=& \; \frac{C^{({\rm Ps})}_{50}}{C^{({\rm Ps})}_{30} \times (10)^2} 
\nonumber\\[1.133ex]
=& \; \left\{ 
\begin{array}{cc} 
0.135 & \;\; \mbox{(conductor)} \\
0.124 & \;\; \mbox{($\alpha$--quartz)} \\
\end{array}
\right.  \,.
\end{align}
A deviation by 13.5\% is larger than the uncertainty 
of many current measurements of the dielectric function
of materials~\cite{Pa1985}.
For hydrogen, the quadrupole correction is 
\begin{align}
\label{ratioH}
\frac{\calE^{({\rm H})}_2(z = 10)}{\calE^{({\rm H})}_1(z = 10)}
=& \; \frac{C^{({\rm H})}_{50}}{C^{({\rm H})}_{30} \times (10)^2}
\nonumber\\[1.133ex]
=& \; \left\{
\begin{array}{cc}
0.0336 & \;\; \mbox{(conductor)} \\
0.0296 & \;\; \mbox{($\alpha$--quartz)} \\
\end{array}
\right.  \,,
\end{align}
which is of the order of a few percent.
These observations are consistent with the 
literature~\cite{TaRa2014}.
In the study of atom-wall interactions,
atomic systems with an exceptionally 
large static polarizability 
have attracted considerable attention.
One example is metastable helium
(in the metastable spin-triplet state)
with a static polarizability of 
315\,a.u. (see Ref.~\cite{CaKlMoZa2005}).
For such systems, one can expect even larger
corrections due to quadrupole effects.
Investigations in these directions are currently
in progress. In the current section, we restrict our attention
to hydrogen and positronium, for which the 
multipole polarizabilities can be 
evaluated in closed analytic form
(see Appendix~\ref{appa}).

%
% Applications to Physisorption
%
\subsection{Applications to Physisorption}
\label{sec4C}

As already outlined above,
the application of van der Waals corrected
density-functional theory to the adsorption
of rare gases on surfaces is a standard
process in surface 
physics~\cite{ZaKo1976,PeBuEr1996,Gr2004,Gr2006,Gr2010,SiSt2008}.
In this context, one adds the van der Waals 
energy which is due to the interaction
with all atoms in the solid, evaluated at the adsorption coordinate,
to a DFT term, which results from the interaction
with the neareast neighbors at the adsorption site.
This approach had been mentioned in the text in 
the upper right column of p.~2280 of Ref.~\cite{ZaKo1976}.
The method has been further developed over a couple of 
decades; the justification for this approach and the general theoretical
background are being 
discussed in Refs.~\cite{ZaKo1976,PeBuEr1996,Gr2004,Gr2006,Gr2010,SiSt2008,%
ChASJo2012,SiAmGrAn2012,TaRa2014}.

One of the most important results
of the current study is the 
additional factor $(\ell + 1)/(2 \ell)$ 
in Eq.~\eqref{genmultipoleH} as compared to
the results communicated in Ref.~\cite{LaDKJe2010pra}.
This correction factor evaluates to $3/4$ for 
the quadrupole term.
This correction factor also affects a few
results recorded in the literature, for example,
the $C_5$ coefficients reported in 
Ref.~\cite{KaSiArSa2022}.
In Ref.~\cite{ZaKo1976}, the authors 
take into account the effect of the 
reference plane at $z_0$ in the modified 
expansion
\begin{equation}
V(z) = - \frac{C_3}{(z-z_0)^3}
- \frac{C'_5}{(z-z_0)^5} - \cdots\,,
\end{equation}
where $C'_5 = C_5 + 6 C_3 z_0^2$, and $C_5$ 
is the quadrupole coefficient from Eq.~\eqref{genmultipoleH},
which reads as follows (in atomic units),
\begin{equation}
C_5 = \frac{3}{16 \pi} \,
\int_0^\infty\dd\omega\, \alpha_2(\ii \omega) \,
\frac{\epsilon(\ii \omega)-1}{\epsilon(\ii \omega)+1}\,.
\end{equation}
Let us consider two examples taken from 
Table~III of Ref.~\cite{TaRa2014},
namely, Kr on Cu(111) and Ar on Pd(111).
For Kr on Cu(111), one has $C_3 = 0.558$ and
$C'_5 = 3.963$ in atomic units
according to Table~III of Ref.~\cite{TaRa2014},
as well as $z_{\rm eq} = 5.99$ (equilibrium position)
and $z_0 = 0.39$. One can easily solve for 
$C_5 = 3.453$ (before correction) and 
$C_5 = 2.590$ (after correction with the 
multiplicative factor $3/4$).
Adding the result from DFT-GGA from 
Table~III of Ref.~\cite{TaRa2014},
which 20.3\,meV, one obtains the 
modified van der Waals corrected adsorption
energy of 121\,meV which is even closer to the 
value of 119\,meV from Ref.~\cite{SiAmGrAn2012}
than the value of 126\,meV given in 
Table~III of Ref.~\cite{TaRa2014}.

For Ar on Pd(111), one has $C_3 = 0.476$ and
$C'_5 = 2.584$ in atomic units
according to Table~III of Ref.~\cite{TaRa2014},
as well as $z_{\rm eq} = 5.59$ (equilibrium position)
and $z_0 = 0.26$. One then solves for 
$C_5 = 2.391$ (before correction) and
$C_5 = 1.793$ (after correction).
Adding the result for DFT-GGA
from Table~III of Ref.~\cite{TaRa2014},
which 14.9\,meV, one obtains the 
modified van der Waals corrected adsorption
energy of 113\,meV which is a bit closer to the
comparison value of 110\,meV from Ref.~\cite{SiSt2008}
than the value of 117\,meV originally given in
Table~III of Ref.~\cite{TaRa2014}.
Further considerations on adsorption energies 
are currently in progress.

%
% Conclusions
%
\section{Conclusions}
\label{sec5}

In this article,
we have (re-)derived (see Ref.~\cite{LaDKJe2010pra})
the quadrupole, octupole and hexadecupole
corrections to the atom-wall interaction,
with a special emphasis on the isolation 
of the relevant angular-momentum components from the derivative
tensors of the electric field.
The functional form of our results is the 
same as the one obtained in
in Ref.~\cite{LaDKJe2010pra},
but important differences are obtained 
for the prefactors.

The general results for the 
dipole, quadrupole, octupole and hexadecupole contributions
to the atom-wall interactions have
been given in Eqs.~\eqref{EEL1res},~\eqref{EEL2res},~\eqref{EEL3res}
and~\eqref{EEL4res}, respectively.
Short-range limits have been given for 
either term in 
Eqs.~\eqref{EEL1short},~\eqref{EEL2short},~\eqref{EEL3short}, 
and~\eqref{EEL4short},
and the long-range limits 
have been analyzed in 
Eqs.~\eqref{EEL1long},~\eqref{EEL2long},~\eqref{EEL3long}
and~\eqref{EEL4long}.
For the long-range limit, we have found rather concise 
formulas for the dependence of the coefficient
multiplying the term proportional 
to $z^{-2\ell-2}$, in terms 
of logarithms [see Eqs.~\eqref{Psi1},~\eqref{Psi2},~\eqref{Psi3}
and~\eqref{Psi4}].
The analytic results for the long-range limit 
constitute an important addition to the 
results originally reported in Refs.~\cite{LaDKJe2010pra}.

Furthermore, we find that the expansion
into multipole terms constitutes an asymptotic,
divergent series [see Eqs.~\eqref{genmultipoleH}
and~\eqref{genmultipolePs}]. Series with factorially 
divergent coefficients can, in many cases,
be summed using generalizations of the 
Borel method~\cite{Je2000prd,CaEtAl2007}.
Furthermore, a truncation of the series
at the smallest term yields an excellent 
approximation to the complete result,
so that the divergent character of the series
is not an obstacle to the deduction of 
theoretical predictions.
We find that it is possible to express
the quadrupole, octupole, and hexadecupole
polarizabilities of hydrogen 
and positronium in closed analytic form (see Appendix~\ref{appa}).
This enables us to reanalyze multipole
corrections, with a special
emphasis on hydrogen and positronium.
We find that the quadrupole correction
is phenomenologically relevant [see Eqs.~\eqref{ratioPs}
and~\eqref{ratioH}]. A concrete application 
is discussed in Sec.~\ref{sec4C}. 
A reanalysis of the dielectric 
function of $\alpha$-quartz using the functional form
given in Eq.~\eqref{new_fit_formula} 
reveals very good agreement with 
numerical data from Ref.~\cite{Ph1985}.
For the short-range coefficients
of the multipole corrections to the atom-surface
interactions (perfect conductor and $\alpha$-quartz),
we present results in Tables~\ref{table2} and~\ref{table3}.
These data confirm the rapid growth of the 
coefficients multiplying the multipole corrections,
both for interactions with perfect conductor and $\alpha$-quartz,
consistent with the eventual factorial divergence
of the series.
Applications to physisorption (van der Waals 
corrected density-functional theory) are discussed
in Sec.~\ref{sec4C}.
The modified result for the quadrupole correction
derived here yields important 
corrections to results previously communicated
in Ref.~\cite{TaRa2014}.

%
% Acknowledgments
%
\section*{Acknowledgments}

The authors acknowledge insightful conversations 
with Carsten A.~Ullrich, Christopher Moore, 
Yakov Itin and Istv\'{a}n N\'{a}ndori.
This research was supported by
NSF grant PHY--2110294.

\appendix

%
% Multipole Polarizabilties of Hydrogen and Positronium
%
\section{Multipole Polarizabilties of Hydrogen and Positronium}
\label{appa}

Hydrogen and positronium constitute 
atomic systems for which the exact 
evaluation of multipole polarizabilities
is possible analytically.
Hence, we focus on these two atomic sytems,
for definiteness, while 
stressing that other atomic systems 
could be more interesting from 
the point of view of applications
(see Sec.~\ref{sec4C}).
In this Appendix, we use atomic units 
(see Chap.~2 of Ref.~\cite{JeAd2022book}) 
with $a_0 = 1$, $E_h = 1$, $\epsilon_0 = 1/(4 \pi)$,
$\hbar = 1$ and $e = 1$ (unit elementary charge).
The atoms of interest are 
hydrogen and positronium.
With techniques outlined in Ref.~\cite{GaCo1970}
and in Chap.~4 
of Ref.~\cite{JeAd2022book},
it is possible to derive closed-form expressions
for hydrogen and positronium
which we use in the nonrelativistic, and non-recoil limit.
We consider the specialization of the
tensor~\eqref{QLm} to a one-electron
atom and work with the matrix element
\begin{multline}
Q_\ell^{({\rm H})}(\omega) =
\frac{1}{2 \ell + 1} \sum_m
\left< Q_{\ell m}
\frac{1}{H - E^{({\rm H})}_{1S} + \omega}
Q^*_{\ell m} \right>_{1S} \,,
\end{multline}
where the superscript indicates the atom
(H stands for hydrogen). From $Q_\ell^{({\rm H})}(\omega)$,
one obtains the $2^\ell$-pole polarizability as
follows,
\begin{equation}
\alpha^{({\rm H})}_\ell(\omega) =
Q_\ell^{({\rm H})}(\omega) +
Q_\ell^{({\rm H})}(-\omega) \,.
\end{equation}
The polarizability of positronium 
can be obtained from the hydrogen result as follows,
\begin{equation}
\alpha^{({\rm Ps})}_\ell(\omega) =
2^{2 \ell + 1 } \left[ 
Q_\ell^{({\rm H})}(2 \omega) +
Q_\ell^{({\rm H})}(-2 \omega) 
\right] \,.
\end{equation}
According to Eq.~(4.154) of Ref.~\cite{JeAd2022book},
the dipole term can be expressed in terms
of a variable $t$,
\begin{equation}
\label{Q1H}
Q_1^{({\rm H})}(\omega) = \frac{2 t^2 \, p^{(1)}(t)}{3 \, (1-t)^5 \, (1+t)^4} 
+ \frac{256 \,t^9 \, f(t)}{3\,(1+t)^5 \,(1-t)^5} \,,
\end{equation}
where the photon energy is parameterized by the $t$ variable,
\begin{equation}
t = t(\omega) = \frac{1}{\sqrt{1 + 2 \omega}} \,.
\end{equation}
The polynomial $p^{(1)}(t)$ incurred in Eq.~\eqref{Q1H}
is 
\begin{equation}
p^{(1)}(t) = 3 - 3 t - 12 t^2 + 12 t^3 + 19 t^4 - 19 t^5 
- 26 t^6 - 38 t^7 \,.
\end{equation}
Within the representation in terms of the $t$ variable,
the transition to positronium amounts to the replacement $t \to t'$
where $t' = t'(\omega) = ( 1 + 4 \omega )^{-1/2}$.
The quadrupole term reads as follows,
\begin{equation}
Q_2^{({\rm H})}(\omega) = 
\frac{t^2 \, p^{(2)}(t)}{5 \, (1-t)^7 \, (1+t)^6} 
- \frac{4096\,t^{11}\,(t^2 - 4) \, f(t)}{5\,(t^2-1)^7} \,.
\end{equation}
The polynomial $p^{(2)}(t)$ reads as follows,
\begin{multline}
p^{(2)}(t) = 
45 - 45 t - 285 t^2 + 285 t^3 + 786 t^4 - 786 t^5 - 1322 t^6 \\
+ 1322 t^7 + 2865 t^8 + 5327 t^9 - 553 t^{10} - 1495 t^{11} \,.
\end{multline}
The octupole term is
\begin{align}
Q_3^{({\rm H})}(\omega) =& \;
\frac{9 \, t^2 \, p^{(3)}(t)}{7 (1-t)^9 (1+t)^8}
\nonumber\\[0.1133ex]
& \; - \frac{ 36864 \, t^{13} \, (t^2 - 4) \, (t^2 - 9) f(t)}{7\,(t^2-1)^9} \,.
\end{align}
It contains the polynomial $p^{(3)}(t)$ which reads as follows,
\begin{multline}
p^{(3)}(t) =
70 - 70 t - 595 t^2 + 595 t^3 + 2280 t^4 - 2280 t^5 - 5309 t^6 
\\
+ 5309 t^7 + 9134 t^8 - 9134 t^9 - 24077 t^{10} - 49651 t^{11}
\\
+ 6532 t^{12} + 20092 t^{13} - 323 t^{14} - 1725 t^{15} \,.
\end{multline}
Finally, the hexadecupole term reads as follows,
\begin{align}
Q_4^{({\rm H})}(\omega) =& \;
\frac{t^2 \, p^{(4)}(t)}{9 (1-t)^{11} (1+t)^{11}}
\nonumber\\[0.1133ex]
& \; - \frac{ 262144 \, t^{15} \, (t^2 - 4) \, (t^2 - 9) \, (t^2 - 16) \, 
f(t)}{9\,(t^2-1)^{11}} \,.
\end{align}
The polynomial $p^{(4)}(t)$ is given by
\begin{multline}
p^{(4)}(t) =
14175 - 14175 t - 150255 t^2 + 150255 t^3 
\\
+ 732060 t^4 
- 732060 t^5 
- 2191500 t^6 + 2191500 t^7 
\\
+ 4624386 t^8 
- 4624386 t^9 - 8049042 t^{10} 
+ 8049042 t^{11} 
\\
+ 23686124 t^{12} + 51811348 t^{13} - 7367996 t^{14} 
\\
- 24613572 t^{15} 
+ 484375 t^{16} + 3316713 t^{17} + 14153 t^{18} 
\\
- 145225 t^{19} \,.
\end{multline}
The asymptotic limits are as follows.
For high photon energy, one obtains the result
\begin{equation}
Q_\ell^{({\rm H})}(\omega) =
\frac{ 2^{2 \ell - 1} \, \Gamma(2 \ell + 3)}{2 \ell + 1} \,
\frac{1}{\omega} 
+ \calO\left( \frac{1}{\omega^2} \right) \,.
\end{equation}
By contrast, the result in the 
static limit is
\begin{equation}
\label{static}
Q_\ell^{({\rm H})}(\omega = 0) =
\frac{ 2 \, (\ell + 2)}{\ell (\ell + 1)} \,
\Gamma(2 \ell + 3) \,,
\end{equation}
for the general multipole order $2^\ell$.
For the static value of the polarizability,
one multiplies the result given in Eq.~\eqref{static}
by a factor $2$.
The result~\eqref{static} can be used in 
Eqs.~\eqref{EEL1long},~\eqref{EEL2long},~\eqref{EEL3long}
and~\eqref{EEL4long}, in order to obtain long-range asymptotics
for a perfect conductor.
For positronium, one multiplies the result given
in Eq.~\eqref{static}
by $2^{2\ell+1}$ in order to obtain 
the static limit of the $2^\ell$-pole polarizability of
positronium.

\end{document}